\documentclass[a4paper,fleqn]{cas-dc}

\usepackage[font=scriptsize, justification=centering]{caption}
\usepackage{subcaption}
\usepackage[switch]{lineno}
\usepackage{multirow}
\usepackage{enumitem}
\usepackage{listings}
\usepackage{xspace}
\usepackage[flushleft]{threeparttable}
\usepackage[numbers]{natbib}
\usepackage{color,soul}
\usepackage{bibnames}
\usepackage{amssymb}
\usepackage{verbatim}
\usepackage{amsmath}
\usepackage{amsthm}
\usepackage{float}
\usepackage{graphicx}
\usepackage{subcaption}
\usepackage[linesnumbered, ruled,vlined]{algorithm2e}
\usepackage{hyperref}
\usepackage{mdframed}
\usepackage{tabularx}
\usepackage{booktabs}
\usepackage{flushend}
\usepackage{mathrsfs}
\usepackage[T1]{fontenc}
\usepackage{balance}

\usepackage{color,soul}
\renewcommand\hl[1]{#1}

\def\tsc#1{\csdef{#1}{\textsc{\lowercase{#1}}\xspace}}
\tsc{WGM}
\tsc{QE}
\tsc{EP}
\tsc{PMS}
\tsc{BEC}
\tsc{DE}

\definecolor{light-gray}{gray}{0.92}  
  {\begin{mdframed}[backgroundcolor=light-gray]\begin{mdtheorem}{name}{label}}%
  {\end{mdtheorem}\end{mdframed}}

\newtheorem{definition}{Definition}[section]

\newcommand{\etal}{\textit{et al.}\space}
\newcommand{\cross}{\textit{cross-project}\xspace}
\newcommand{\dev}{\textit{dev.-process}\xspace}
\newcommand{\tool}{\textsc{PatchExplainer}\xspace}

\begin{document}
\let\WriteBookmarks\relax
\def\floatpagepagefraction{1}
\def\textpagefraction{.001}

\shorttitle{\tool}

\shortauthors{Vu \textit{et~al.}}

\title [mode = title]{Automated Description Generation for Software Patches}

\author{Thanh Trong Vu}[orcid=0009-0008-3377-6565]
\ead{thanhvu@vnu.edu.vn}

\affiliation{organization={Faculty of Information Technology, VNU University of Engineering and Technology},
    city={Hanoi},
    country={Vietnam}}

\author{Tuan-Dung Bui}[orcid=0000-0001-7136-7529]
\ead{21020006@vnu.edu.vn}

\author{Thanh-Dat Do}
\ead{20020045@vnu.edu.vn}

\author{Thu-Trang Nguyen}[orcid=0000-0002-3596-2352]
\ead{trang.nguyen@vnu.edu.vn}

\author{Hieu Dinh Vo}[orcid=0000-0002-9407-1971]
\ead{hieuvd@vnu.edu.vn}

\author{Son Nguyen}[orcid=0000-0002-8970-9870]
\ead{sonnguyen@vnu.edu.vn}
\cormark[1]

\cortext[cor1]{Corresponding author}


\begin{abstract}
Software patches are pivotal in refining and evolving codebases, addressing bugs, vulnerabilities, and optimizations. Patch descriptions provide detailed accounts of changes, aiding comprehension and collaboration among developers. However, manual description creation poses challenges in terms of time consumption and variations in quality and detail. In this paper, we propose \tool, an approach that addresses these challenges by framing patch description generation as a machine translation task. In \tool, we leverage explicit representations of critical elements, historical context, and syntactic conventions. Moreover, the translation model in \tool is designed with an awareness of description similarity. 
Particularly, the model is \textit{explicitly} trained to recognize and incorporate similarities present in patch descriptions clustered into groups, improving its ability to generate accurate and consistent descriptions across similar patches. 
The dual objectives maximize similarity and accurately predict affiliating groups. Our experimental results on a large dataset of real-world software patches show that \tool consistently outperforms existing methods, with improvements up to 189\% in \textit{BLEU}, 5.7X in \textit{Exact Match} rate, and 154\% in \textit{Semantic Similarity}, affirming its effectiveness in generating software patch descriptions.

\end{abstract}

\begin{keywords}
Software Patch, Patch Description, Neural Machine Translation, Code-to-Text
\end{keywords}



\maketitle

\section{Introduction}

Software patches~\cite{real-patches,patch-research-2,understanding-patch} are crucial components that facilitate the ongoing refinement and evolution of codebases. These patches are fundamental in addressing various issues within a software system, ranging from addressing bugs and vulnerabilities to optimizations. 
Patch descriptions play a vital role in articulating the nature and purpose of these modifications, providing a detailed account of the changes introduced, the specific issues addressed, and the methodologies employed in the patches~\cite{SECOM}. 
\textbf{}This descriptive layer not only serves as a documentation tool but also fosters effective communication and collaboration among developers~\cite{good_commit_message,SECOM}. By explaining the `\textit{what},' `\textit{where}', and `\textit{how}' of code changes, patch descriptions contribute to a comprehensive understanding of the evolving software, enabling developers to navigate through the details of modifications, collaborate seamlessly, and ensure the stability and functionality of the software system throughout its lifecycle.

Manual creation of patch descriptions poses inherent challenges in the software development cycle. The details involved in translating technical code changes into precise, coherent, and informative descriptions present a time-consuming task for developers. This process not only consumes time but also introduces the potential for inconsistencies and variations in the quality of descriptions across different developers. Moreover, the diverse backgrounds and perspectives of developers may lead to discrepancies in the level of detail and clarity provided in descriptions. 
Thus, automating patch description generation is an essential yet highly challenging task. 
Indeed, automated description generation approaches face the following major challenges:

\textbf{Contextual Understanding of Code Changes:} A significant challenge in patch description generation is the understanding of contextual information within code changes. The approaches must accurately capture the specific details of bug fixes introduced in the code. The complicated relationships between code elements, as well as the bug addressed and the operations employed to address it could pose challenges for the approaches to comprehend and express accurately in natural language.

\textbf{Syntactic and Stylistic Consistency:} Ensuring syntactic and stylistic consistency in generated patch descriptions is another major challenge. The approaches must capture the language preferences of individual developers and specific projects. 
The diversity in language preferences and the limitations in understanding the broader software patching context could affect the description generation approaches' ability to replicate specific language choices accurately.

Conceptually, describing software patches can be viewed as a machine translation (MT) task, where the patch code is the source language and the corresponding description is the target language.
Recently, MT techniques, particularly neural machine translation (NMT) techniques, have been increasingly applied across various domains in software engineering~\cite{codesum_paper,icse20,deep-comment-gen, Logentext,nngen,coregen,race,come,bug-explainer}. NMT has found utility in tasks such as code summarization~\cite{codesum_paper,icse20}, code comment/log generation~\cite{deep-comment-gen, Logentext}, and commit message generation~\cite{nngen,coregen,race,come}. 
Standard NMT models comprise two components\cite{nmt,gg_nmt}: an encoder and a decoder. The encoder's role is to initially comprehend the meaning of words in the source language and convert them into an intermediate numeric representation. After that, the decoder generates target words sequentially, drawing on the intermediate representation and preceding words from the generated sequence.
In an ideal scenario, NMT models are anticipated to accurately grasp the meaning of an input sequence and subsequently articulate it in the output sequence.
The objective of generating similar descriptions for similar software patches is \textit{implicitly} expected to be achieved through this process of encoding and decoding within the NMT model.
%
%
%

In this paper, we introduce \tool, a novel automated approach that tackles the challenges in patch description generation. 
To enhance contextual understanding, our approach is to incorporate explicit representations of critical elements within patch descriptions, conveying information about bugs/vulnerabilities, operational fixing details, and patching scope. 
Additionally, we consider historical software patching context to employ syntactic and stylistic conventions and enforce consistency. 

Furthermore, to foster accuracy in generating patch descriptions by MT models, our approach is to design a patch description generation model as a translation model with the awareness of the similarity between descriptions.
Particularly, the model is \textit{explicitly} trained to recognize and incorporate similarities present in various patch descriptions, improving its ability to generate accurate and consistent descriptions across similar patches.
The intuition is that instilling the model with the ability to additionally discern the similarity between descriptions can contribute to more precise description generation.
In particular, we handle the similarity between the descriptions by clustering the patches with similar descriptions into a group.
To implement the idea of a translation model with the awareness of descriptions' similarity, our approach leverages dual objectives: one is the machine translation task's original objective, \textit{correctly generating expected descriptions}, and the other is \textit{accurately predicting the affiliating group containing similar patches reflected in the descriptions}.

We conducted experiments to evaluate the effectiveness of \tool for automated patch description generation in comparison to the state-of-the-art approaches~\cite{nngen,coregen,race, come}. 
%
%
Our results show that \tool is consistently better than existing methods across different metrics such as \textit{BLEU}, \textit{Exact Match} rate, and \textit{Semantic Similarity}, demonstrating significant improvements in linguistic precision, semantic alignment, and overall coherence in patch descriptions. 
Particularly, \tool significantly outperforms the existing approaches, with relative improvements ranging from \textbf{43\%} to \textbf{189\%} in \textit{BLEU}. The \textit{Exact Match} rate of our approach surpasses that of all compared methodologies by \textbf{5.7X}, emphasizing its proficiency in generating descriptions that precisely match reference texts. 
Furthermore, \tool achieves considerably better \textit{Semantic Similarity} compared to the state-of-the-art approaches, with improvements ranging from \textbf{16\%} to an impressive \textbf{154\%}, affirming its ability to capture the underlying meaning or intent of patches. The consistent and substantial improvements observed across diverse metrics emphasize the robustness and effectiveness of our approach in automating the patch description generation process. 

The contributions of this paper are listed as follows:

\textbf{1. Contextual Representation}: A novel method that uses both semantic and conventional contexts for automated patch description generation.

\textbf{2. Novel technique}: \tool, a novel automated patch description generation approach, is developed based on machine translation with dual objectives, capturing both semantic and conventional information to generate patch description.

\textbf{3. Extensive experimental results}: An extensive experimental evaluation showing the performance of {\tool} over the state-of-the-art methods for patch description generation.

\section{Motivating example}

\subsection{Examples}
\label{sec:example}

This section illustrates the task of automated description generation for software patches via an example. 

\begin{figure}
    \centering
    \includegraphics[width=0.9\columnwidth]
    {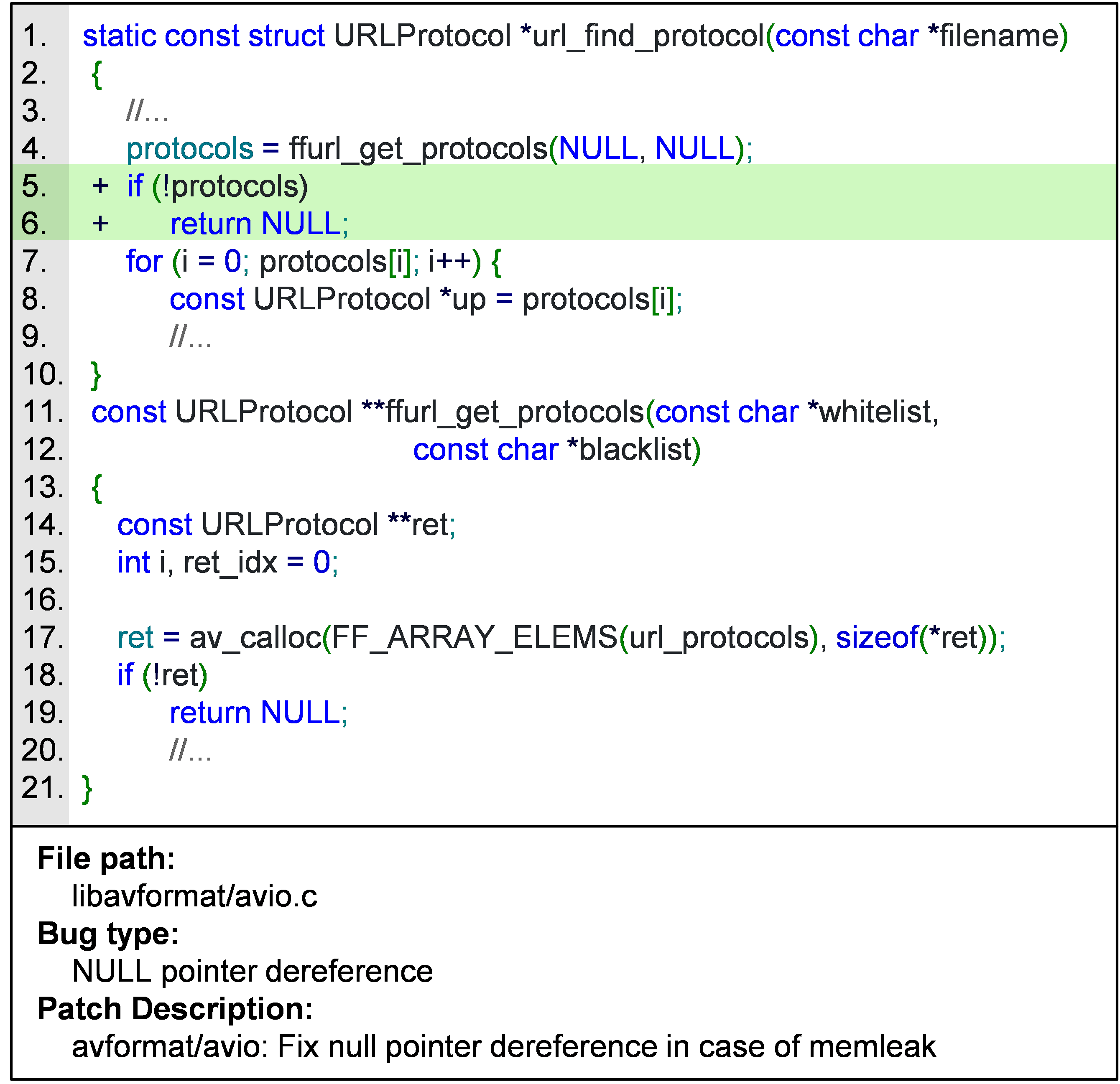}
    \caption{A patch in project \texttt{FFmpeg} and its description}
    \label{fig:example}
\end{figure}


%
%
Figure~\ref{fig:example} shows a patch in project \texttt{FFmpeg}\footnote{https://github.com/FFmpeg/FFmpeg/commit/936751b} and its corresponding description. 
In this example, function \texttt{url\_find\_protocol} finds and returns a suitable URL protocol for a given \texttt{filename}. However, this function could encounter a NULL pointer dereference problem (line 7) if the list of protocols, \texttt{protocols}, returned by function \texttt{ffurl\_get\_protocols} is null (line 4). To avoid this problem, an \texttt{if} statement was added (lines 5--6) to prevent this function from accessing a null \texttt{protocols} list. 
The author of the commit, \textit{Michael Niedermayer}, described the patch as \textit{``avformat/avio: Fix null pointer dereference in case of memleak''}.
%


The description provides descriptive information about the patch: \textit{\textbf{what}}, \textit{\textbf{where}}, and \textit{\textbf{how}} the patch is fixing.
In Figure~\ref{fig:example}, the patch was introduced to fix a \textit{NULL pointer dereference} issue in \texttt{avio.c} of module \texttt{avioformat} by preventing the use of null \texttt{protocols} to handle \textit{memory leak} cases.
%
To understand better the \textit{what}, \textit{where}, and \textit{how} aspects in the real-world patch descriptions, we performed a rule-based analysis, which has been used in the existing studies on change to text~\cite{type-ist,mark_study}, on 24.2K patches and their corresponding description collected from the datasets published by the existing studies~\cite{wang2021PatchDB}. 
Additionally, we manually analyzed a set of randomly sampled patches to carefully investigate these aspects and minimize the threats to the validity of the rule-based analysis.
Specifically, for the description $d$ of a patch $p$, we applied the rule-based analysis to determine if $d$ provides a certain aspect:
\begin{enumerate}
   
    \item To determine whether the description contains the information about \textit{what} the patch is fixing bugs, we investigate whether the information about bug types is encoded in the description. Particularly if $d$ contains certain keywords about various types of bugs~\cite{type-ist}, such as \textit{`mem', `null', `leak', `overflow', 
 `injection', `typo',  `bound', `crash', `unused'}, or \textit{`uninit'}, then $d$ provides the information about what issues $p$ is fixing, the \textit{what} aspect of $p$,
 
    \item To determine whether the description contains information about how the patch is fixing bugs, we investigate whether the information about code modifications or actions is included. If $d$ contains verbs or modification patterns describing the changes to fix the buggy version~\cite{mark_study}, such as \textit{`add', `fix', `check', `remove', `delete', `avoid', `prevent', `reject'}, \textit{`replace'}, or \textit{`Add [missing] ... [for|to...]`} then $d$ provides the information about how $p$ is fixing issues, the \textit{how} aspect of $p$, and

    \item If $d$ starts with the name of functions/files/modules where the code modifications of $p$ are made, then $d$ provides the information about where $p$ is fixing issues, the \textit{where} aspect of $d$.
\end{enumerate}
%

In our empirical study, we found that 11.0\% of the descriptions contain all three aspects, while 57.6\% of patch descriptions express 2/3 aspects, and about 89.0\% of them convey at least one of the three aspects. 
Among the set of patches that convey at least one of the three aspects, we randomly sampled 385 cases with a confidence level of 95\% and margin of error of 5\% to investigate whether these patch descriptions contain information other than the \textit{what}, \textit{how}, and \textit{where} aspects. After manually investigating the sampled cases, we found that only 7.8\% of them contained other kinds of information, such as when the issue occurred.
This implies that the majority of patch descriptions focus primarily on the essential aspects of \textit{what}, \textit{how}, and \textit{where}, with a limited proportion including supplementary details. 

Moreover, among 385 patches randomly sampled from the remaining 11\% of the patches whose descriptions do not contain the aforementioned keywords, we observed that they still provide information about \textit{what}, \textit{how}, and \textit{where} yet in different ways.
For example, \textit{``Reduce picture size for yadif.''}\footnote{https://github.com/FFmpeg/FFmpeg/commit/02d0803}.
More details about our empirical analysis can be found on our website~\cite{website}.

As seen, the patch description in Figure~\ref{fig:example} is written following the convention, \textit{``\{where\}: \{what and how the patch fixes the bug\}''}. This convention is \textit{consistently} applied by the author (\textit{Michael Niedermayer}) in describing more than 90\% of his patches in \texttt{FFmpeg}. 
For example, two descriptions of the other patches created by the same developer in \texttt{FFmpeg} before the patch in Figure~\ref{fig:example} are \textit{``avformat/avio: fix memory leak in url\_find\_protocol''} and \textit{``avformat/avio: Fix unknown protocol handling''}.
The recent study~\cite{history_paper} has shown that syntactic and stylistic conventions of a particular project or user are important in describing code changes.

Overall, \textit{the descriptions of patches provide descriptive information about \textbf{what}, \textbf{where}, and \textbf{how} the patch is fixing \textbf{in personalized or project-specific natural language}}.

\subsection{Key ideas}

To enhance the effectiveness of patch description generation, our approach revolves around three core ideas: explicitly providing contextual information, enforcing syntactic and stylistic conventions through historical context, and framing patch description generation as a translation task.

\textit{1. Explicitly Providing Contextual Information for Patch Description Generation}. 
Our first key idea focuses on the explicit provision of contextual information to convey essential elements in patch descriptions. Specifically, we aim to offer detailed insights into the \textit{what}, \textit{how}, and \textit{where} aspects of a patch addressing a bug. This involves identifying and presenting information about the bug being addressed (\textit{what}), detailing the operations employed to fix the bug (\textit{how}), and specifying the locations (e.g., functions, files, or modules) where modifications occur (\textit{where}). By incorporating these aspects, our model learns to utilize such information effectively, enabling the generation of informative and context-rich descriptions for software patches.

\textit{2. Enforcing Syntactic and Stylistic Conventions Through Historical Context}. 
The second key idea focuses on enforcing syntactic and stylistic conventions within a specific project or for an individual developer. By considering the historical context of patches, our model tailors its output to align seamlessly with established coding standards and preferences. This is achieved by incorporating recent patch descriptions as an input of our model, allowing the model to learn from the syntactic and stylistic conventions observed in the project's or developer's past patches. This approach ensures that the generated patch descriptions not only adhere to linguistic norms but also reflect individual or project-specific coding conventions, enhancing the integration of the generated patches into the existing codebase.

\textit{3. Patch Description Generation as Machine Translation with Dual-objective}.
Our third key idea is to frame patch description generation as a machine translation task with dual objectives.
Besides the traditional objective in NMT, correctly generating expected descriptions, we also enforce NMT models to \textit{explicitly} recognize and incorporate the similarities between descriptions.
Particularly, the additional objective is for each patch, accurately predicting the affiliating group containing similar patches reflected by the descriptions. 
This third idea involves learning to express the \textit{what}, \textit{how}, and \textit{where} aspects of a patch in personalized or project-specific natural language and to determine the group to which the patch belongs, facilitating a smoother comprehension of the introduced code changes. 

\section{\tool: Automated Description Generation for Software Patches}


In this section, we introduce \tool, a novel automated description generation for software patches. In \tool, there are two main phases: \textit{Contextual Information Extraction} and \textit{Patch Description Generation}.

\subsection{Contextual Information Extraction}
A software patch is a structured representation of changes applied to a code repository. In this work, we formally define software patches as follows.

\begin{definition}{\textbf{(Software Patch)}}. Given a repository, a software patch $p$ is a 5-tuple $p=\langle C, \Delta, a, t, d \rangle$ where:
    \begin{itemize}
    \item $C$ is a buggy code version
    \item $\Delta = \{c_1, c_2, ..., c_n\}$ is a set of code changes. A change $c \in \Delta$ is a pair $c = \langle o, s \rangle$ where: 
        \begin{itemize}
        \item $o$ is an operation change, $o \in \{added,deleted\}$
        \item $s$ is a changed statement
        \end{itemize}
    \item $a$ is the patch author
    \item $t$ is the fixing time
    \item $d$ is the patch description
    \end{itemize} 
\end{definition}{}

\subsubsection{Semantic Context Extraction}

For a patch, code changes are introduced to fix a bug.
In other words, the aspect of \textit{how} the code is corrected is conveyed by the code changes.
The changes impact the (existing) buggy code to correct the code's unexpected behaviors.
Intuitively, the buggy version should be required to capture \textit{what} bug in the code is fixed by the patch.
However, not every part of the buggy code is semantically related to the bug and patch. To avoid redundancy in capturing the meaning of the patch, the code statements unrelated to the bug and patch should not be provided. 
For example, in Figure~\ref{fig:example}, the code that is not related to the changed statements at lines 5 and 6 via program dependencies (i.e., control/data dependency), such as lines 7--9, should not be considered. 
In this work, we represent the information conveying the \textit{what} and \textit{how} aspects of a patch in a compact representation (\textit{Patch-related Code}) containing the patch's code changes and their related (unchanged) code of the buggy code version. 
%
%
%
%

\begin{definition}{\textbf{(Patch-related Code)}}. For a patch $p=\langle C, \Delta, a, t, d \rangle$, the patch-related code of $p$, $R_p$ is the code sequence containing (annotated) changed statements in $\Delta$ and the unchanged statements in $C$ which are \textbf{semantically} related to the changed statements. 
Specifically, the set of related unchanged statements, $U$, consists of statements in $C$ and not in $\Delta$, such that $s \in U$ if: 
    \begin{itemize}
        \item $\exists c = \langle o, s' \rangle \in \Delta$, such that there exists a data/control dependency between statement $s$ and statement $s'$, or 
        \item $\exists s'' \in U$, such there exists a data/control dependency between the statements $s$ and $s''$.
    \end{itemize}
\end{definition}{}
Note that the statements in the patch-related code of $p$ remain the same relative order in the buggy code version $C$ because the order of the statements is crucial for determining the meaning of the code.


The \textit{scope} of a patch provides information about \textit{where} a specific code segment is changed, i.e.,  changed function or file. Thus, we will leverage the information related to the file and function that contains the changed code in a patch. In our empirical study, we found that about 73\% of the patch descriptions contain the changed functions/files' names.
%

\begin{definition}{\textbf{(Patch Scope)}}. For a patch $p = \langle C, \Delta, a, t, d \rangle$, the patch scope of $p$, $S_p$, consists of the information about the file names and function names where $p$ modifies in the buggy code $C$.
\end{definition}{}

For example, the scope of the patch shown in Fig \ref{fig:example} includes the name of the changed function, \texttt{url\_find\_protocol}, and the path of the modified file, 	\textit{``libavformat/avio.c''}. In this example, the file path is used at the beginning of the patch description.

\subsubsection{Syntactic Context Extraction}

\textit{The syntactic and stylistic conventions} in patches' descriptions help patches and their descriptions readability and searchability. These conventions refer to the rules that order/group the information about patches (including the \textit{what}, \textit{how}, and \textit{where} aspects) by a certain style. However, the conventions of patches could be very different for each project and individual developers~\cite{history_paper}. 
To aid syntactic and stylistic conventions of a particular project or developer in patch description generation, we consider their historical patch descriptions as a part of our generation model's input.

\begin{definition}{\textbf{(Historical Descriptions)}}. For a patch $p = \langle C, \Delta, a, t, d \rangle$, the set of historical descriptions of $p$, $H_p$, consists of the patch descriptions which are created by the same author $a$ of $p$ before fixing time $t$ in the same project.
\end{definition}{}

For instance, the description of the patch shown in Figure~\ref{fig:example} is \textit{``avformat/avio: Fix null pointer dereference in case of memleak''}, while a few historical descriptions of the patch are \{\textit{``avformat/avio: fix memory leak in url\_find\_protocol''} and \textit{``avformat/avio: Fix unknown protocol handling''}\}. All of these descriptions start with the information in the scope of the corresponding patches.

In this work, we formulate the task of patch description generation as a translation task transferring a sentence, $\mathcal{S}_p$ in the source language to the corresponding sentence, $\mathcal{T}_p$ in the target language. In this translation task, $\mathcal{S}_p$ from the source language is the sequence constructed from a software patch $p = \langle C, \Delta, a, t, d \rangle$, $\mathcal{S}_p = concat[R_p, concat(S_p), concat(H_p)]$, where $concat$ is the concatenating function returning a sequence joining sub-sequences by separators; $R_p$, $S_p$, and $H_p$ are the patch-related code, patch scope, and historical description set of $p$. Meanwhile, $\mathcal{T}_p$ from the target language is the description $d$ of $p$. 

\subsection{Patch Description Generation}

\begin{figure*}
 \centering
 \includegraphics[width=2\columnwidth]{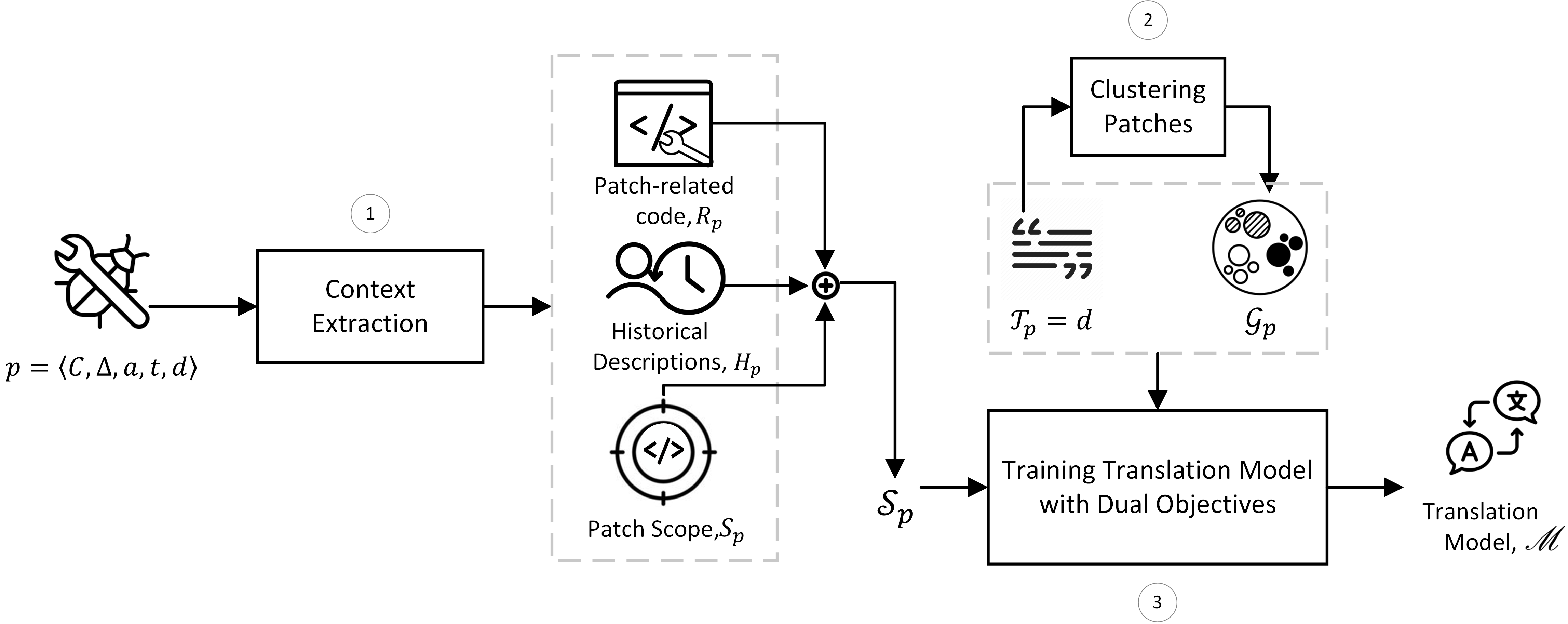}
 \caption{\tool: An Automated Patch Description Generation Approach}
 \label{fig:generation}
\end{figure*}

Our idea is to design a patch description generation model as a translation model with the awareness of the similarity between descriptions.
The rationale is that enforcing the model learns to additionally the similarity between descriptions could help the model generate descriptions more accurately. 
The similarity between patch descriptions could be manifested in various forms to improve patch description generation, such as recurring bug-fix types, shared keywords or phrases, and similar severity levels of addressed issues. 
%
However, the challenge lies in the absence of explicit labels indicating such similarities between patches and their descriptions. To capture implicit/hidden similarities between descriptions, we introduce a crucial step in our approach: \textit{Clustering patches based on description similarity} (Step 2 in Figure~\ref{fig:generation}). This step enables us to organize patches into groups based on shared characteristics and identify meaningful patterns within the varied range of patch descriptions.
%
%
%
%
%

Subsequently, our translation model is trained with a dual-objective: \textit{accurately generating descriptions for individual patches} and \textit{precisely predicting the cluster or group to which each patch belongs} (Step 3 in Figure~\ref{fig:generation}). This dual-task framework leverages the benefits of both semantic grouping and accurate description generation, enabling the model to capture the meaning of code changes while fostering coherence and consistency within identified clusters. 
Moreover, predicting cluster affiliation could be considered an easier ``short-term'' task than the harder ``long-term'' task, patch description generation. The existing studies~\cite{standley2020tasks} have shown that the model tends to perform better once it is trained with both the ``short-term'' and ``long-term'' objectives.
The impact of the consideration similarity between patch descriptions on the description generation performance of this dual-task framework will be experimentally evaluated in Section~\ref{sec:results}.

\subsubsection{Clustering Patches Based on Description Similarity}

Our approach incorporates a crucial step, clustering patches based on the similarity of their descriptions. This process creates cohesive groups of patches with shared semantic characteristics in their language representations. By employing established algorithms~\cite{clustering_survey}, such as $K$-means or hierarchical clustering, we categorize patches into distinct groups where descriptions exhibit notable similarity.

For each patch $p = \langle C, \Delta, a, t, d \rangle$ in the training data, the description $d$ is transformed into $D$-dimensional hidden features, $v_p$ using embedding techniques~\cite{embedding_emse22}, such as TF-IDF~\cite{tfidf}, Word2vec~\cite{word2vec_1}, or CodeBERT~\cite{codebert}.
Next, the clustering process relies on standard algorithms~\cite{clustering_survey} that assess the semantic similarity of patch descriptions. These algorithms use the embedded vectors corresponding to the patch descriptions to categorize patches into groups with similar descriptions. Formally, based on vector $v_p$, patch $p$ is categorized into the group $\mathcal{G}_p$ containing the patches whose descriptions are semantically similar to $d$ in this step.
The primary goal is to capture detailed semantics in descriptions, identifying patches that address similar characteristics, such as types of issues, or relate to specific functionalities within the codebase.
The grouped patches serve as enriched training data for subsequent phases, contributing to the creation of a robust translation model. 

\subsubsection{Training Translation Model with Dual Objectives}

Following patch clustering based on description similarity, a translation model, $\mathscr{M}$, is applied to learn to \textit{generate descriptions} and \textit{predict cluster affiliation} from the provided contextual information.
Particularly, for each patch $p = \langle C, \Delta, a, t, d \rangle$ in the training data, $\mathscr{M}$ is trained to learn generating the description ($\mathcal{T}_p = d$) and predicting cluster affiliation of $p$ (the clustered group $\mathcal{G}_p$). 
%
%
$\mathscr{M}$ is trained with dual-objective to enhance the accuracy and coherence of generated descriptions.

\textit{Objective 1, Accurate Description Generation}:
The first objective involves training the model to generate descriptions for individual patches accurately. This task is central to ensuring that the model captures the meaning of the provided contextual information, including \textit{patch-related code}, \textit{patch scope}, and \textit{historical descriptions}, and communicates them effectively in natural language. 
Specifically, we compare the accuracy of the predicted translation ($\hat{\mathcal{T}}_p$) to the actual translation ($\mathcal{T}_p$) of the source sentence $\mathcal{S}_p$ to compute a loss. While there are several varieties of loss functions, we apply the following common version of the Cross-Entropy Loss:
\begin{equation}
\label{L1}
\mathcal{L}_1 = -\sum^{|\hat{\mathcal{T}}_p|}_{w = 1} \sum^{|V|}_{e = 1} y_{w,e} \log(\hat{y}_{w,e})
\end{equation}

In Formulation~\ref{L1}, $V$ is the list of vocabulary entries. Additionally, $y_{w,e} = 1$ if the vocabulary entry $e$ is the corrected word; otherwise, $y_{w,e} = 0$. This means that if the $w^{th}$ word of $\mathcal{T}_p$ is the vocabulary entry $e$ in $V$, then $y_{w,e} = 1$; otherwise, $y_{w,e} = 0$. Meanwhile, $\hat{y}_{w,e}$ is the predicted probability of vocabulary entry $e$ on word $w^{th}$. 
In essence, $\mathcal{L}_1$ sums over the negative log likelihoods that the model gives to the correct word at each position in $\hat{\mathcal{T}}_p$. Given that the negative log function returns $0$ when the input is $1$ and increases exponentially as the input approaches $0$, the closer the probability that the model gives to the correct word at each point in $\hat{\mathcal{T}}_p$ is to 100\%, the lower the loss.

\textit{Objective 2, Cluster Affiliation Prediction}:
Simultaneously, the second objective entails training the model to predict each patch's cluster or group affiliation. This additional task encourages the model to understand broader patterns and themes within the dataset, fostering coherence and consistency in the generated descriptions. 
For this objective, we compare the accuracy of the predicted cluster affiliation ($\hat{\mathcal{G}}_p$), given $\mathcal{S}_p$, to the cluster affiliation ($\mathcal{G}_p$) determined by the patch clustering step (Step 2 in Figure~\ref{fig:generation}). Formally:

\begin{equation}
\label{L2}
\mathcal{L}_2 = -\sum^{N}_{i = 1} g_{i} \log(\hat{g}_{i})
\end{equation}

In Formulation~\ref{L2}, $N$ is the number of groups/clusters. Additionally, $g_{i} = 1$ if the patch is clustered into the group $i^{th}$; otherwise, $g_{i} = 0$. Meanwhile, $\hat{g}_{i}$ is the predicted probability that the patch in the group $i^{th}$.
This can be done by considering the group index as a special token in target sequences.

\textit{Dual-Task Framework}:
The dual-task framework leverages the benefits of both accurate description generation and cluster affiliation prediction. By optimizing the model for these two objectives, we aim to strike a balance between precision in individual descriptions and a broader understanding of semantic clusters. Thus, the overall loss that the training process has to optimize for each software patch in the training data is $\mathcal{L} = \frac{\mathcal{L}_1 + \mathcal{L}_2}{2}$.
\section{Evaluation Methodology}
\label{sec:eval}
To evaluate our automated description generation approach for software patches, we seek to answer the following research questions:

\noindent\textbf{RQ1: \textit{Accuracy and Comparison}.} How accurate is {\tool} in generating software patch descriptions? And how is it compared to the state-of-the-art approaches~\cite{nngen, codedisum, coregen, race, come}?

\noindent\textbf{RQ2: \textit{Intrinsic Analysis}.} How do different aspects of our approach impact the description generation performance of \tool, including clustering algorithms and the description generation model's properties/components? 

\noindent\textbf{RQ3: \textit{Context Analysis}.} How do provided contextual information, including patch scopes and historical descriptions impact {\tool}'s performance?

\noindent\textbf{RQ4: \textit{Sensitivity Analysis}.} How do various input's factors, including training data size and changed code's complexity, and bug types affect {\tool}'s performance?

\noindent\textbf{RQ5: \textit{Time Complexity}.} What is  {\tool}'s running time?

\subsection{Dataset}

\begin{table}
\centering
\caption{Dataset statistics}\label{tab:dataset}
\resizebox{\columnwidth}{!}{%
\begin{tabular}{lrrrrr}\toprule
\textbf{Repo} &\textbf{\#Patches} &\textbf{AVG. tokens in $d$} &\textbf{\#Changed files} &\textbf{\#Changed LOCs} \\\midrule
Linux &7,877 &7.36 &8,570 &42,190 \\
Ffmpeg &5,092 &7.30 &5,314 &24,708 \\
Media-tree &3,571 &7.32 &3,921 &19,146 \\
Qemu &3,039 &6.82 &3,349 &17,706 \\\midrule
\multicolumn{5}{c}{550 projects more...} \\\midrule
Total &30,308 &7.33 &33,092 &163,277 \\
\bottomrule
\end{tabular}
}
\end{table}

To evaluate patch description generation performance, we construct a large dataset of 30K software patches fixing bugs and vulnerabilities in 554 open-source projects (Table~\ref{tab:dataset}). This dataset is collected from other public datasets in the existing work~\cite{wang2021PatchDB,bug-explainer} and National Vulnerability Database (NVD)~\cite{nvd}.


%
%
To extract the patch descriptions, we applied the same procedure as the method applied by Mahbub \etal~\cite{bug-explainer}. Particularly, the commit messages of these patching commits could be considered as their descriptions. 
We also followed the same data cleaning step of those studies to remove unique tokens (e.g., commit IDs, URLs, developer names, bug IDs, emails), which could be impossible to generate correctly. 
Additionally, we found that 96\% of the collected patches have descriptions containing 3 to 15 tokens.
Thus, we eliminated the cases where the descriptions were not representative, were too short (fewer than three tokens), or were too long (more than 15 tokens). 

\subsection{Evaluation Setup, Procedure, and Metrics}

\subsubsection{Empirical Procedure} 

\textbf{RQ1. Accuracy and Comparison}. 

\textbf{\textit{Baselines.}} We compared \tool against the state-of-the-art patch description generation approaches: 

\begin{enumerate}
 \item \textbf{NNGen}~\cite{nngen} is an IR-based commit message prediction technique.
 
 \item \textbf{Coregen}~\cite{coregen} is a pure Transformer-based approach for representation learning targeting commit message generation.

 \item \textbf{RACE}~\cite{race} is a context-aware retrieval-based deep commit message generation approach combining language modeling and retrieval advantages.
 
 \item \textbf{COME}~\cite{come} combines retrieval techniques with translation-based methods through a decision algorithm, and this method learns better contextualized code change representation

 \item \textbf{GPT-3.5 Turbo}~\cite{gpt} is a large language model (LLM) capable of generating different creative text formats, including descriptions of code changes in the task of patch description generation. 
\end{enumerate}
%
%
%
We used the implementation in their original papers for all the baseline approaches.

\textit{\textbf{Data Splitting Strategies.}} In this comparative study, to assess the generalizability of the approaches, we evaluated the performance of the approaches in two real-world settings: \textit{development-process} and \textit{cross-project}:

\textit{Development- (dev.-) process setting}. 
%
This strategy investigates how well the models perform on unseen data from a different time period.
Particularly, we divided the patches into those before and after time point $t$. The patches before $t$ were used for training, while the patches after $t$ were used for evaluation. We selected a time point $t$ to achieve a training/testing split ratio of 80/20 based on time. Specifically, for the \textit{dev.-process} setting, the patches from Aug 1987 to Sep 2016 were used for training, and the patches from Oct 2016 to Jul 2023 were used for evaluation. In total, the training/testing split in the number of patches for this setting is 24,264/6,062.

\textit{Cross-project setting}. We evaluated how well the approaches can learn to generate patch descriptions in one set of projects and generate descriptions for patches in the other. 
Specifically, the patches in a fixed set of projects were used to train the approaches, and the remaining patches were used for testing. 
For this setting, the whole set of projects was randomly split into 80\% (455 projects) for training and 20\% (107 projects) for testing. The training/testing split in the number of patches for this setting is 24,375/5,952.

These two splitting strategies allow us to evaluate the model's performance in different scenarios:
\begin{itemize}
    \item \textit{Dev.-process} setting (\textit{time-aware} evaluation) assesses the model's ability to adapt to potential changes in coding styles or bug patterns over time.
    \item \textit{Cross-project} setting assesses the model's ability to generalize to unseen projects, demonstrating its broader applicability.
\end{itemize}

\textit{\textbf{Evaluation Setup.}} 
For a practical evaluation, we considered a maximum of ten statements in the patch-related code and the last ten historical descriptions of each patch.
%
%
We used the $K$-mean clustering algorithm and Code Llama 7B~\cite{codellama} parameters fine-tuned using Lora~\cite{hu2021lora} as the translation model in \tool. 
%
For detailed implementation and other hyper-parameters, one can see our website~\cite{website}.
Note that all our experiments were run on a server with GeForce RTX 4090 GPU. 


\textbf{RQ2. Intrinsic Analysis}. We also investigated the impact of different aspects of our
approach on the description generation performance:

\begin{itemize}
 \item Description generation model: We investigated the impact of the description generation model's properties on the performance of \tool, including the translation model and the additional objective (\textit{Cluster Affiliation Prediction}).
 
 \item Patch clustering: We studied the impact of various aspects of the clustering step, including the clustering algorithm and embedding techniques on \tool's performance.

\end{itemize}

\textbf{RQ3. Context Analysis}. We investigated the impact of the provided contextual information on the description generation performance of \tool. Note that, the patch-related code is considered the essential part of patches. Thus, we evaluated the impact of providing path scope and historical descriptions as input of \tool. 

\textbf{RQ4. Sensitivity Analysis}. We studied the impacts of the following factors on the performance of \tool: training size and change complexity. To systematically vary these factors, we gradually added more training data and varied the range of the change rate.

\subsubsection{Metrics}
To evaluate the patch descriptions generated by \tool, we used several metrics at both \textit{surface} and \textit{semantic} levels. In general, the surface-level metrics, including \textit{BLEU}, \textit{METEOR}, \textit{ROUGE}, and \textit{Exact Match}, focus on lexical and syntactic aspects of descriptions, while the semantic-level metrics delve into the meaning or semantics of the generated content. Collectively, these metrics provide a comprehensive evaluation of the quality and alignment of generated descriptions with reference descriptions in the patch description generation task.

1) \textit{Surface-Level Metrics}:

\begin{itemize}
 \item \textbf{BLEU} (\textit{Bi-Lingual Evaluation of Understanding}) measures how many word sequences from the actual description occur in the generated description and uses a $n$-gram precision to generate a score:
 $$
 BLEU = BP \cdot \exp(\sum_{n = 1}^{N} w_n \log(p_n))
 $$
 where $p_n$ is the ratio between overlapping $n$-grams from both generated and actual descriptions, and the total number of $n$-grams in the generated description, and $w_n$ is the weight of the $n$-gram length. Following the existing studies~\cite{come,race,nngen,coregen}, we use $N = 4$ and $w_n = 0.25$ for all $n \in [1, N]$. The brevity penalty, $BP$, lowers \textit{BLEU} if the generated one is too small.
 
 \item \textbf{METEOR} (\textit{Metric for Evaluation of Translation with Explicit ORdering}) is an F-Score-Oriented metric for measuring the performance of translation models, $MET. = F\text{-weighted}_{M}(1-p)$, where $p$ is the chunk penalty, $p=0.5(c/u_m)^3$, with $C$ is the number of chunks in the generated description, $U_m$ is the number of unigrams in the candidate. The harmonic mean of unigram precision and recall, with recall weighted higher than precision, $P_M = m/w_t$, $R_M = m/w_r$, and $F\text{-weighted}_{M} = \frac{10P_M R_M}{R_M + 9P_M}$, where $m$ is a number of unigrams in the generated description also found in the actual one, while $w_t$ and $w_r$ are the numbers of unigrams in generated one and actual one.
 
 \item \textbf{ROUGE} (\textit{Recall-Oriented Understudy for Gisting Evaluation}) is one set of metrics for comparing the automatically generated descriptions against the actual descriptions. We focus on \textit{ROUGE-L}, which computes the Longest Common Subsequence ($LCS$). Particularly, \textit{ROUGE-L} recall:
 $$R_{RL} = \frac{num\_tokens(LCS(\mathcal{T}, \hat{\mathcal{T}}))}{k}$$ 
 and \textit{ROUGE-L} precision:
 $$P_{RL} = \frac{num\_tokens(LCS(\mathcal{T}, \hat{\mathcal{T}}))}{t}$$
 where $k$ and $t$ are the numbers of tokens in $\mathcal{T}$ and $\hat{\mathcal{T}}$; \textit{ROUGE-L} F1-score $F\text{-}1_{RL} = \frac{2P_{RL} R_{RL}}{R_{RL} + P_{RL}}$.
 
 \item \textbf{Exact Match} measures the rate of the cases where the generated description exactly matches the corresponding actual description. It is analogous to string equality checks in many programming languages, which are case-sensitive and space-sensitive.
\end{itemize}

2) \textit{Semantic-level Metric}: In this work, we applied a similar semantic-level metric utilizing deep-pretrained word embedding techniques to measure the semantic similarity between the generated description and the actual one. Following the existing studies~\cite{bug-explainer,haque2022semantic}, we apply Sentence-BERT~\cite{sentence-bert}, trained on a large amount of code and text to provide a fixed-length numeric representation for any given description. We compute the \textit{Semantic Similarity} as follows: 
$$ SemSim(\mathcal{T}_p, \hat{\mathcal{T}}_p) = cosine(sbert(\mathcal{T}_p), sbert( \hat{\mathcal{T}}_p))$$
where $sbert(x)$ is the numerical representation from Sentence-BERT for any input text $x$, $\mathcal{T}_p$ is the actual description, and $\hat{\mathcal{T}}_p$ is the generated description.

Note that, for every considering metric $M$, a description generation approach with higher $M$ is better.

\section{Experimental Results}
\label{sec:results}

\subsection{Answering RQ1: Accuracy and Comparison}

\begin{table}
\centering
\caption{\textit{Cross-project:} Patch description generation performance}
\label{tab:comparison.cross} 
\resizebox{\columnwidth}{!}{%
\begin{tabular}{lrrrrrr}\toprule
\textbf{Method} &\textbf{\textit{BLEU} } &\textbf{\textit{MET}. } &\textbf{ROU.} &\textbf{\textit{EM}} &\textbf{\textit{SemSim}} \\\midrule
NNgen &9.8 &6.65 &8.13 &0.49 &25.02 \\
CoreGen &11.12 &7.93 &13.51 &0.02 &33.29 \\
GPT-3.5 Turbo &13.50 &13.50 &12.85 &0.07 &50.30 \\
RACE &18.23 &15.61 &19.26 &0.29 &51.69 \\
COME &18.56 &16.66 &19.44 &0.57 &53.25 \\

\tool &\textbf{23.94} &\textbf{20.53} &\textbf{27.06} &\textbf{1.26} &\textbf{58.69} \\
\bottomrule
\end{tabular}
}
\end{table}

\begin{table}
\centering
\caption{\textit{Dev.-process:} Patch description generation performance}
\label{tab:comparison.dev} 
\resizebox{\columnwidth}{!}{%
\begin{tabular}{lrrrrrrr}\toprule
&\textbf{\textit{BLEU}} &\textbf{\textit{MET.}} &\textbf{\textit{ROU.}}  & \textbf{\textit{EM}} & \textbf{\textit{SemSim}} \\\midrule
NNgen &8.98 &6.03 &6.92  &0.00 &23.93 \\
CoreGen &10.86 &7.21 &11.99  &0.05 &30.02 \\
GPT-3.5 Turbo &13.92 &13.86 &13.68 &0.02 &50.94 \\
RACE &18.48 &15.53 &18.21 &0.00 &49.10 \\
COME &18.20 &16.09 &18.82  &0.16 &52.39 \\
\tool &\textbf{25.98} &\textbf{22.01} &\textbf{28.99}  &\textbf{1.40} &\textbf{60.78} \\
\bottomrule
\end{tabular}
}
\end{table}

Table~\ref{tab:comparison.cross} and Table~\ref{tab:comparison.dev} show the performance of \tool and the existing approaches, including NNgen~\cite{nngen}, CoreGen~\cite{coregen}, RACE~\cite{race}, COME~\cite{come}, and GPT-3.5 Turbo in the \cross and \dev settings. 
As seen, \tool significantly outperforms all the state-of-the-art approaches in all considering metrics in both settings.

For both settings, in terms of \textit{surface-level} similarity, \tool demonstrates substantial performance enhancements when compared to COME. For instance, the relative improvements of \tool compared to COME in the \dev setting are significant: \textbf{43\%} in \textit{BLEU}, \textbf{37\%} in \textit{METEOR}, and \textbf{54\%} in \textit{ROUGE-L F1}. Meanwhile, \tool's performance is \textbf{over three to four folds} better than those of NNGen, CoreGen, and RACE across all these metrics. \tool consistently outperforms the existing approaches, these signify a significant advancement in linguistic precision.
Especially, \tool's \textit{Exact Match} rate surpasses those of NNGen, CoreGen, RACE, and COME, reflecting its ability to generate descriptions that precisely match the reference descriptions.
In terms of \textit{semantic-level} similarity, \tool consistently raises the bar, demonstrating improvements ranging from \textbf{10\%} to an impressive \textbf{154\%} when compared to the existing approaches. This indicates a substantial advancement in capturing the underlying meaning or intent of code changes.

Moreover, while the performances of the other approaches remain stable between the two settings, \tool's performance could improve by 9\% in the \dev setting compared to its performance in the \cross setting. The reason for this improvement could be that \tool could effectively learn and utilize the vocabulary in the same project's past.

Overall, \textit{the consistent and significant improvements across these metrics demonstrate \tool's effectiveness in not only achieving high linguistic precision but also in conveying richer semantic content, demonstrating the high potential of our technique in generating patch descriptions}.

\begin{figure}
    \centering
    \includegraphics[width=\columnwidth]
    {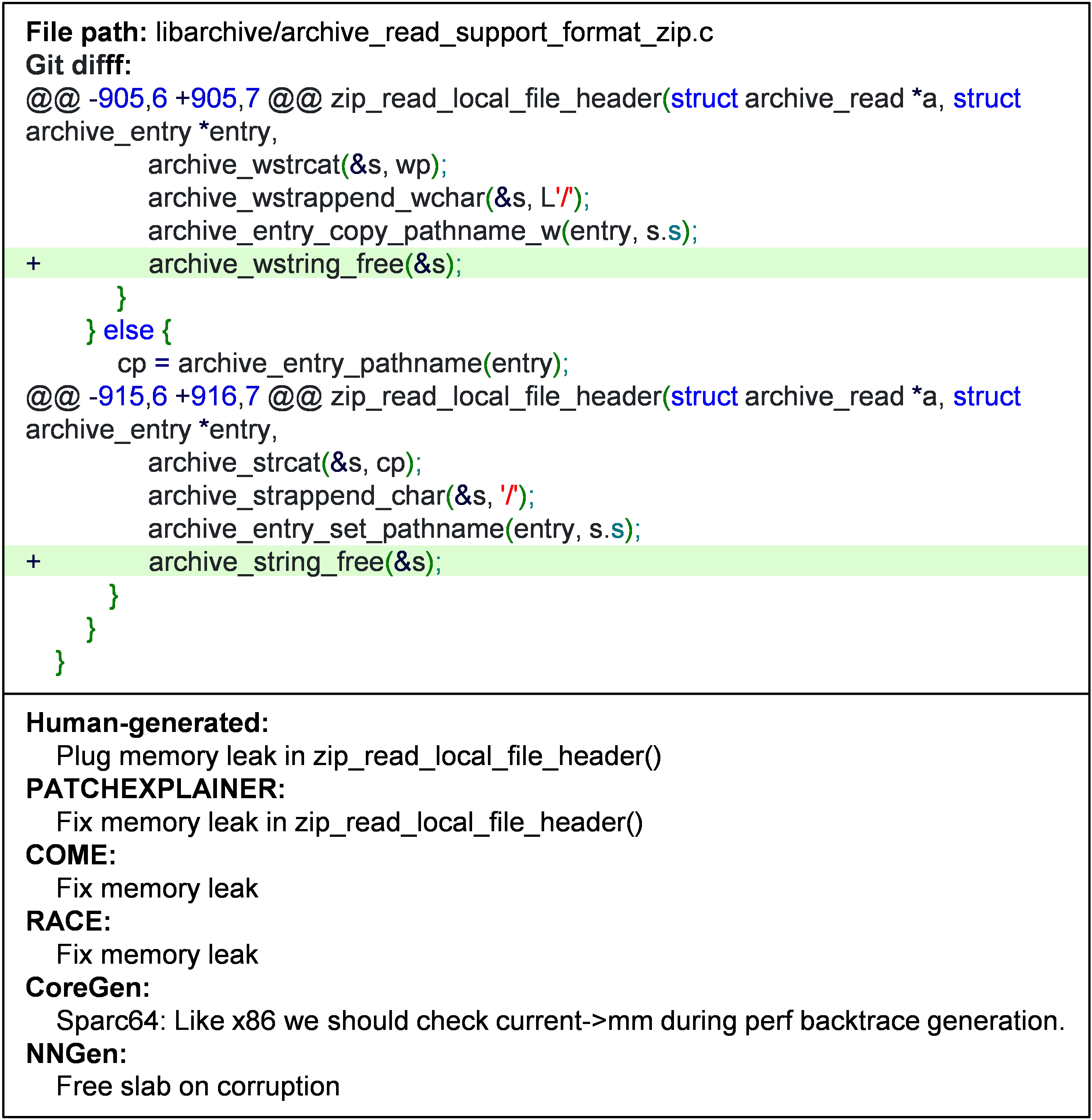}
    \caption{The descriptions generated by \tool and the others for patch \texttt{8fd7839} in \texttt{libarchive}}
    \label{fig:good_example}
\end{figure}

\textbf{\textit{Result Analysis}}.
Analyzing the results, we found that the main reason for the higher performance of \tool compared to the existing approaches is that \tool provides explicit context to generate patch descriptions. Moreover, by training the translation model with dual-objective, the generated descriptions contain precise and comprehensive semantics in multiple aspects. 
For example, Figure \ref{fig:good_example} shows a patch in \texttt{libarchive}\footnote{https://github.com/libarchive/libarchive/commit/8fd7839}. The description generated by \tool correctly contains the \textit{what} (\texttt{"memory\ leak"}) and \textit{where} (function \texttt{"zip\_read\_local\_file\_header"}) which match the information in the ground truth. 
On the other hand, the descriptions generated by COME and RACE can only provide the bug type fixed by the patch, while the descriptions generated by CoreGen and NNGen are completely irrelevant. 

\begin{figure}
    \centering
    \includegraphics[width=\columnwidth]
    {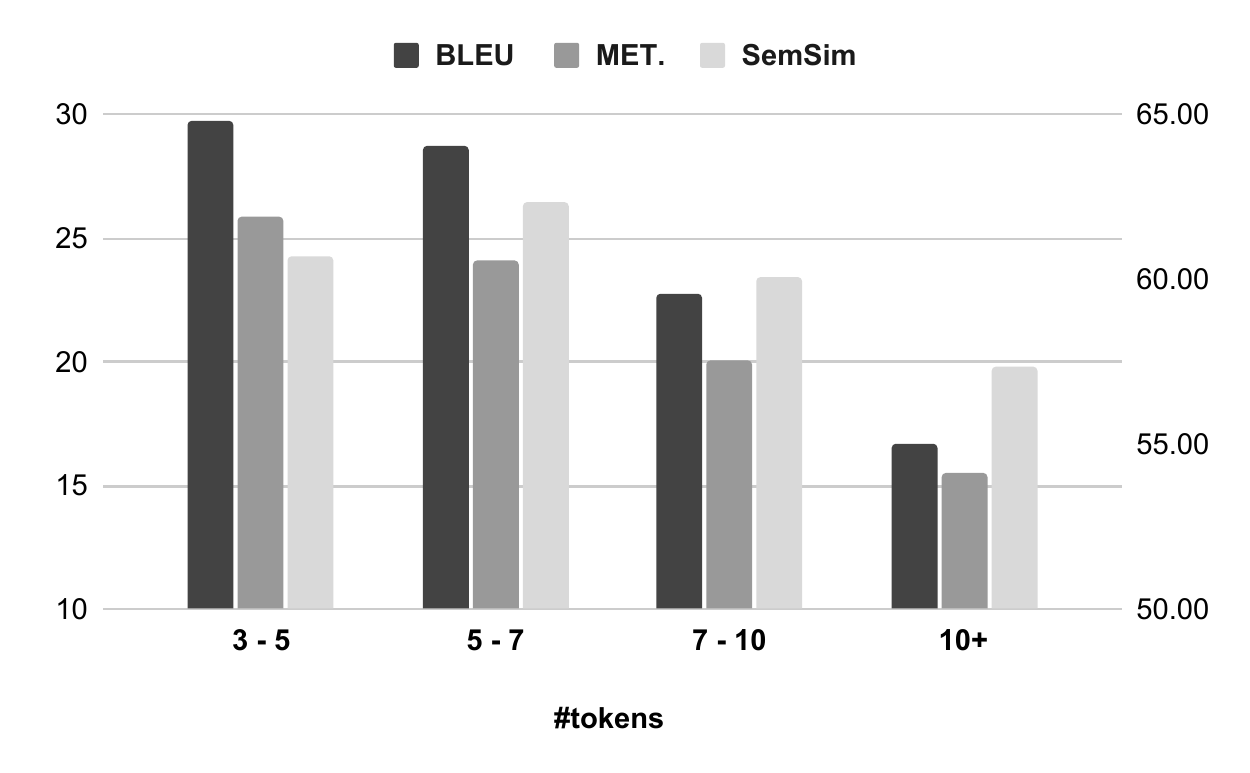}
    \caption{Performance of \tool in different 
 description complexity levels (left axis: \textit{BLEU} \& \textit{MET}.; right axis: \textit{SemSim})}
    \label{fig:description_complexity}
\end{figure}

\begin{figure}
    \centering
    \includegraphics[width=\columnwidth]
    {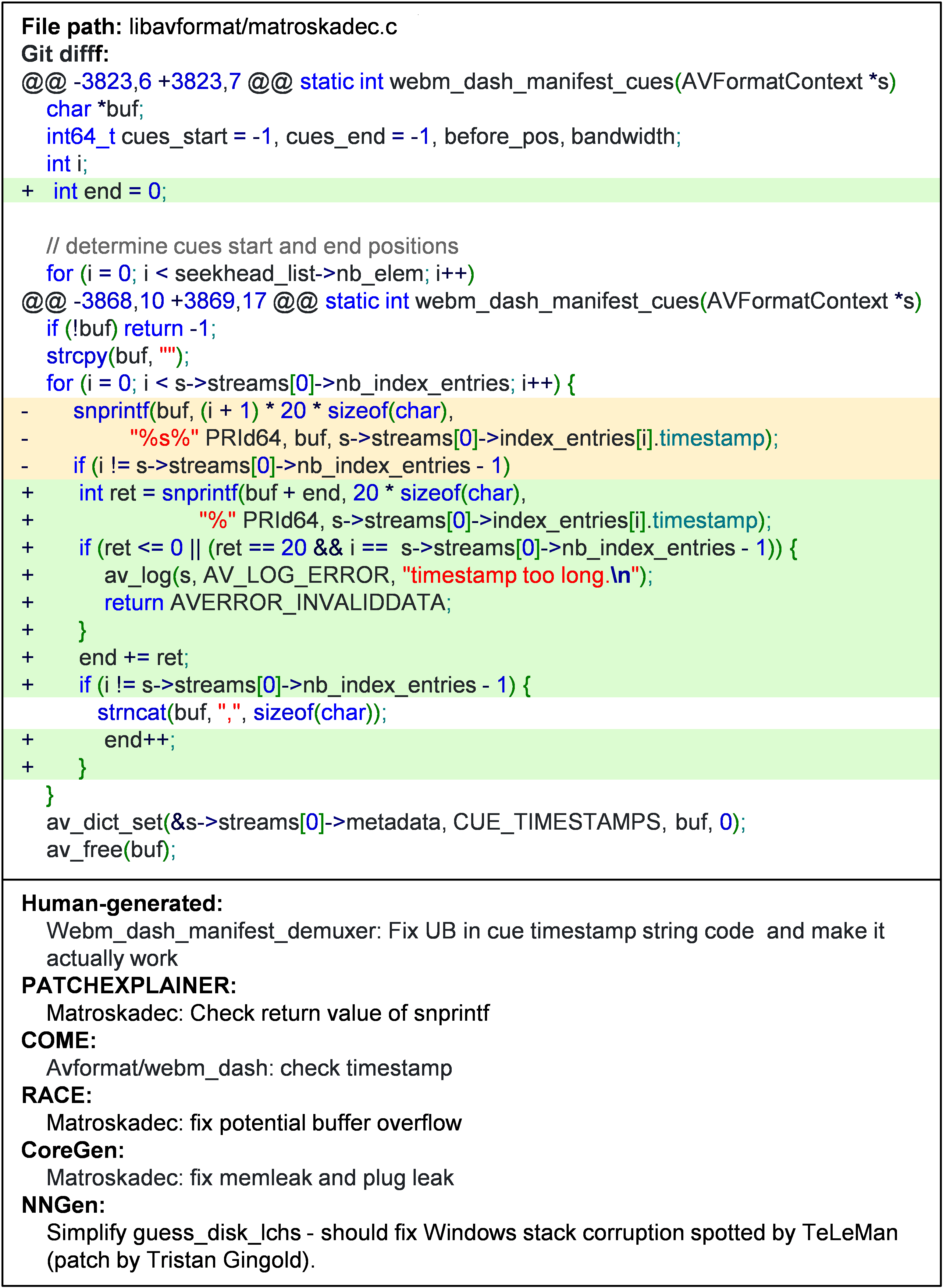}
    \caption{Patch \texttt{8e6b9ef} in \texttt{FFmpeg} and the descriptions}
    \label{fig:bad.example}
\end{figure}

Among the cases where \tool performed poorly, our observation is that a large portion of them have long descriptions. This is expected because it is harder to generate longer descriptions. Indeed, as shown in Figure~\ref{fig:description_complexity}, the performance of \tool decreases gracefully when generating longer descriptions. 
Figure~\ref{fig:bad.example} shows low-quality descriptions generated by \tool and the other approaches. In this example, even though the description generated by \tool is quite different from the expected one, it still correctly expresses information about the changed file and operational fixing details. In contrast, the other techniques fail to deliver accurate information about the patch.
%
%

\begin{figure}
    \centering
    \includegraphics[width=\columnwidth]
    {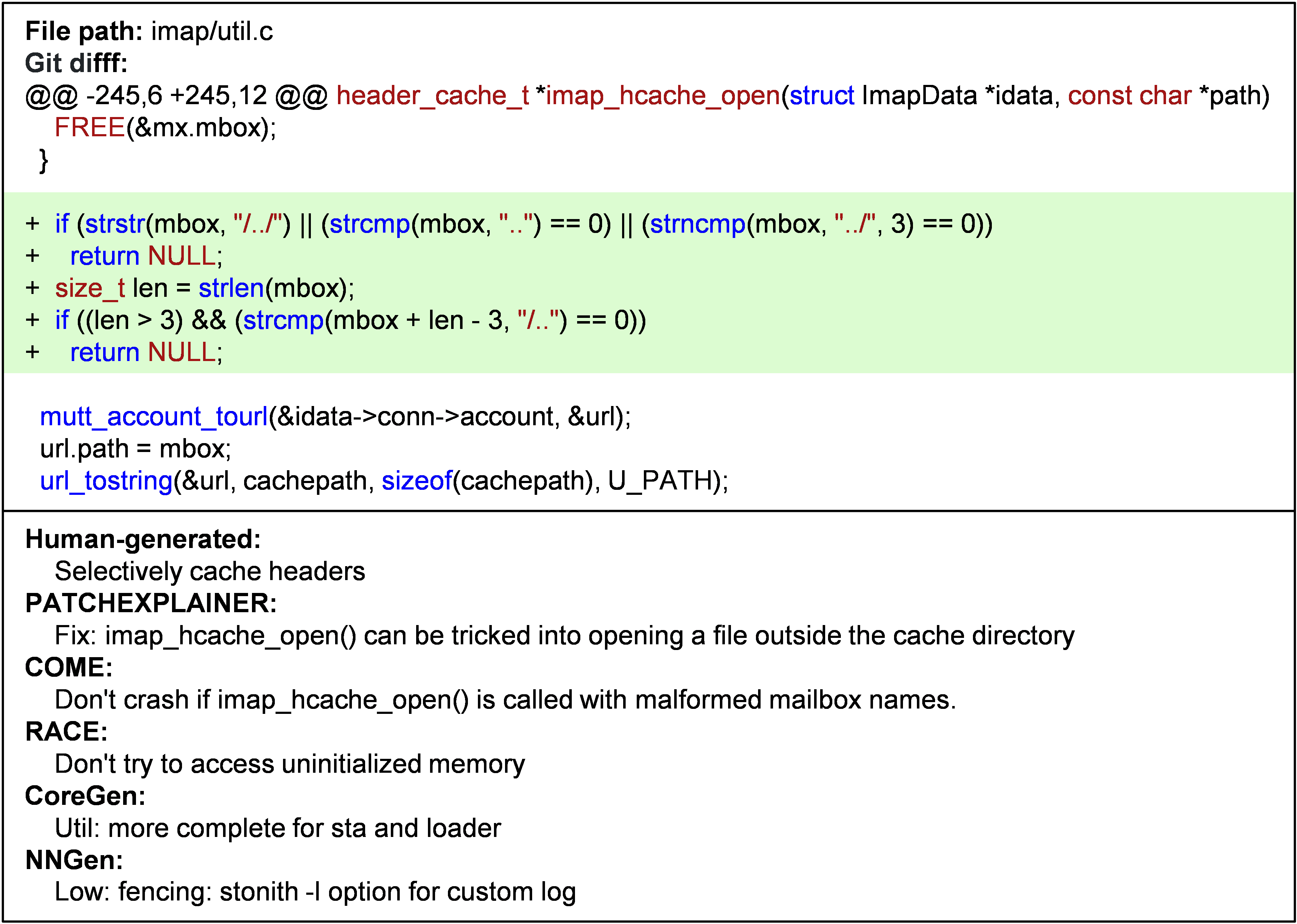}
    \caption{An interesting generated patch description by \tool (CVE-2018-14355)}
    \label{fig:interesting-example}
\end{figure}

\textbf{\textit{Interestingly}}, we found several cases where the descriptions were not well generated by \tool regarding the considered metrics. Yet, they were even more informative than the human-generated ones. For example, the patch in Figure~\ref{fig:interesting-example} addresses mishandling \texttt{``..''} directory traversal in a mailbox name in function \texttt{imap\_hcache\_open}\footnote{https://nvd.nist.gov/vuln/detail/CVE-2018-14355}.
The description written by the developer for this patch is not very informative. COME can generate an acceptable description for this case. Meanwhile, the description generated by \tool correctly describes the bug fixed by the patch and the fixed function.

To further validate \tool's generalizability, we conducted an additional evaluation on the MCMD dataset (C/C++)~\cite{mcmd}, which is commonly used to benchmark commit message generation approaches~\cite{race,come}. As shown in Table~\ref{tab:mdmc}. \tool achieves significant improvements across all evaluated metrics, with the improvements of 10--68\% in \textit{BLEU}, 17--81\% in \textit{METEOR}, and 20--79\% in \textit{ROUGE-L}. These results on the MCMD dataset further support the effectiveness of \tool.  
%

\begin{table}
\centering
\caption{Performance comparison on MCMD dataset~\cite{mcmd} (C/C++)}
\label{tab:mdmc}
\begin{tabular}{lrrrr}\toprule
&\textbf{BLEU} &\textbf{MET. } &\textbf{ROU.} \\\midrule
NNgen &13.61 &9.39 &18.21 \\
CoreGen &16.74 &11.72 &22.83 \\
RACE &19.13 &12.55 &24.52 \\
COME &20.80 &14.55 &27.01 \\
\tool &22.88 &17.03 &32.66 \\
\bottomrule
\end{tabular}
\end{table}



\subsection{Answering RQ2: Intrinsic Analysis}

\subsubsection{Impact of Translation Model}

\begin{table}
\centering
\caption{Performance of \tool by translation model}
\label{tab:trans_model}
\resizebox{\columnwidth}{!}{%
\begin{tabular}{lrrrrrrr}\toprule
&\textbf{\textit{BLEU}} &\textbf{\textit{MET.}} &\textbf{\textit{ROU.}}  & \textbf{\textit{EM}} & \textbf{\textit{SemSim}} \\\midrule
Transformer &11.86	&7.62	&9.57	&0.08	&29.45 \\
CodeT5 &24.23 &21.93 &25.78 &1.17 &59.29 \\
Code Llama &25.98 &22.01 &28.99  &1.40 &60.78 \\
\bottomrule
\end{tabular}
}
\end{table}

To evaluate the impact of translation models on the performance of \tool, we compare the performance of \tool with three representative translation models: Transformer~\cite{vaswani2017attention}, CodeT5~\cite{codet5}, and Code Llama 7B~\cite{codellama} in the \dev setting. 
In this experiment, we use all of the contexts of \tool; we use the $K$-means algorithm with \textit{Tf–idf} embedding to cluster description and apply it to the \dev setting. 
The performance of \tool with different translation models is shown in Table~\ref{tab:trans_model}. As expected, \tool achieves the best performance with Code Llama 7B, which can capture the meaning of code and text~\cite{codellama}. Meanwhile, \tool with CodeT5 with 220M parameters shows quite competitive results, with a margin of 7\% and a 119\% improvement in BLEU compared to \tool with Code Llama and traditional Transformer (110M parameters). The reason is that CodeT5 and Code Llama are larger models and pre-trained with much data. However, CodeT5 and Code LLama cost 2--10 times in training and inferring. Thus, the transformer could be suitable to be applied in \tool when deploying on machines with limited resources. At the same time, CodeT5 and Code Llama could be suitable for the cases preferring the patch description generation effectiveness.

\subsubsection{Impact of Training with Dual-Objective}

Table~\ref{tab:multi-objective} compares the performance of \tool under different training variants: Single-Objective and Dual-Objective.
The results in Table~\ref{tab:multi-objective} demonstrate that the Multi-Objective training variant outperforms the Single-Objective variant across all metrics. Particularly, the Multi-Objective training yields an improvement of 6.3\% and 3.5\% in \textit{BLEU} and \textit{Semantic Similarity}, respectively. Especially, the Exact Match rate of \tool with Dual-Objective is about 80\% higher than that of \tool with Single-Objective.
%
%
%
These results suggest that incorporating the dual objectives of maximizing similarity between generated and actual descriptions and accurately predicting affiliating groups significantly enhances the overall performance of \tool. 
Therefore, the dual-objective training approach is recommended for optimal automated patch description generation results.

\begin{table}
\centering
\caption{Performance of \tool by training variants}
\label{tab:multi-objective}
\scriptsize
\resizebox{\columnwidth}{!}{%
\begin{tabular}{lrrrrrr}\toprule
&\textbf{\textit{BLEU}} &\textbf{\textit{MET}. } &\textbf{\textit{ROU}.} &\textbf{\textit{EM}} &\textbf{\textit{SemSim}} \\\midrule
Single-Objective &24.45 &20.49 &27.52 &0.79 &58.75 \\
Dual-Objective &25.98 &22.01 &28.99 &1.40 &60.78 \\
\bottomrule
\end{tabular}
}
\end{table}

\subsubsection{Impact of Clustering Algorithm}

In Table~\ref{tab:clustering_algorithm}, we present the evaluation of \tool's performance using different representative clustering algorithms~\cite{clustering_survey}. The considered algorithms include $K$-means (Centroid-based), DBScan (Density-based), Gaussian Mixture (Distribution-based), and Agglomerative (Hierarchical), each configured with an identical number of clusters.
The results show the stability of \tool across diverse clustering algorithms, showcasing consistent performance with maximum variations of 1.9\% and 0.5\% in \textit{BLEU} and \textit{Semantic Similarity}, respectively. This robust performance shows that \tool's effectiveness is maintained regardless of the clustering algorithm employed. Thus, the choice of clustering algorithm should lean towards simplicity and efficiency, making lightweight algorithms like $K$-means more preferable over sophisticated alternatives such as Gaussian Mixture.

We further investigated the influence of different embedding methods on the clustering performance of \tool, particularly when using the $K$-means algorithm. The considered embedding techniques include Tf-idf~\cite{tfidf}, Word2Vec for non-contextual embedding~\cite{word2vec_1}, and CodeBERT for contextual embedding~\cite{codebert}—all representative methods from prior work~\cite{embedding_emse22}.
Table \ref{tab:embedding} illustrates the stability of \tool's performance across various embedding methodologies during the clustering step. Notably, \tool showcases consistent and reliable results regardless of the chosen embedding method. This indicates that \tool's efficacy in generating patch descriptions remains robust across different ways of representing input sequences.
Among the evaluated embedding methods, even the traditional and interpretable Tf-idf approach proves to be a viable choice for \tool. This observation suggests that simplicity and interpretability, as offered by Tf-idf, can be favored without compromising the effectiveness of \tool in the context of patch description generation.

In general, with the common objective of grouping similar patches of the studied clustering algorithms, the groups resulting from the different algorithms and embedding techniques still contain the similar patches within them. Thus, \tool with dual-objectives training can effectively recognize and incorporate the similarities between patches in generating descriptions.

\begin{table}
\centering
\caption{Performance of \tool by clustering algorithms}
\label{tab:clustering_algorithm}
\resizebox{\columnwidth}{!}{%
\begin{tabular}{lrrrrrr}\toprule
\textbf{} &\textbf{\textit{BLEU}} &\textbf{\textit{MET.}} &\textbf{\textit{ROU.}}  & \textbf{\textit{EM}} & \textbf{\textit{SemSim}} \\\midrule
DBScan &24.32 &21.96 &25.87 &1.25 &59.52 \\
$K$-mean &24.23 &21.93 &25.78 &0.98 &59.29 \\
Gaussian Mixture &24.15 &21.83 &25.66 &1.22 &59.24 \\
Agglomerative &24.60 &22.15 &26.04 &1.47 &59.52 \\
\bottomrule
\end{tabular}
}
\end{table}

\begin{table}
\centering
\caption{Performance of \tool by embedding methods}
\label{tab:embedding}
\resizebox{\columnwidth}{!}{%
\begin{tabular}{lrrrrrr}\toprule
\textbf{} &\textbf{\textit{BLEU}} &\textbf{\textit{MET.}} &\textbf{\textit{ROU.}}  & \textbf{\textit{EM}} & \textbf{\textit{SemSim}} \\\midrule
Tf-idf &24.23 &21.93 &25.78 &0.98 &59.29 \\
Word2vec &24.43 &22.01 &25.85 &1.35 &59.40 \\
CodeBERT &24.34 &21.90 &25.71 &1.29 &59.45 \\
\bottomrule
\end{tabular}
}
\end{table}

\subsection{Answering RQ3: Context Analysis}


\subsubsection{Impact of Patch Scope}
To study the impact of considering the patch scope on \tool's performance, we compare the performance of \tool using the inputs with and without the patch scopes. In this experiment, we use CodeT5 as a translation model, using the $K$-means algorithm with \textit{Tf–idf} embedding to cluster descriptions. 
As shown in Table ~\ref{tab:patch_scope}, the performance of \tool increases by 33\% in \textit{BLEU} and 17\% in \textit{Semantic Similarity} when considering the patch scope as a part of the input for generating description. 
This result aligns well with our empirical about the \textit{where} aspect of patch descriptions in practice and confirms our method considering patch scopes as a part of models' input. 
%

\begin{table}
\centering
\caption{Impact of patch scope on \tool's performance}
\label{tab:patch_scope}
\resizebox{\columnwidth}{!}{%
\begin{tabular}{lrrrrrrr}\toprule
\textbf{} &\textbf{\textit{BLEU}} &\textbf{\textit{MET.}} &\textbf{\textit{ROU.}}  & \textbf{\textit{EM}} & \textbf{\textit{SemSim}} \\\midrule
W/O Patch Scope &18.26 &16.86 &20.45  &0.36 &50.67 \\
W/ Patch Scope &24.23 &21.93 &25.78 &0.98  &59.29 \\
\bottomrule
\end{tabular}
}
\end{table}

\subsubsection{Impact of Historical Descriptions}
\begin{figure}
    \centering
    \includegraphics[width=\columnwidth]
    {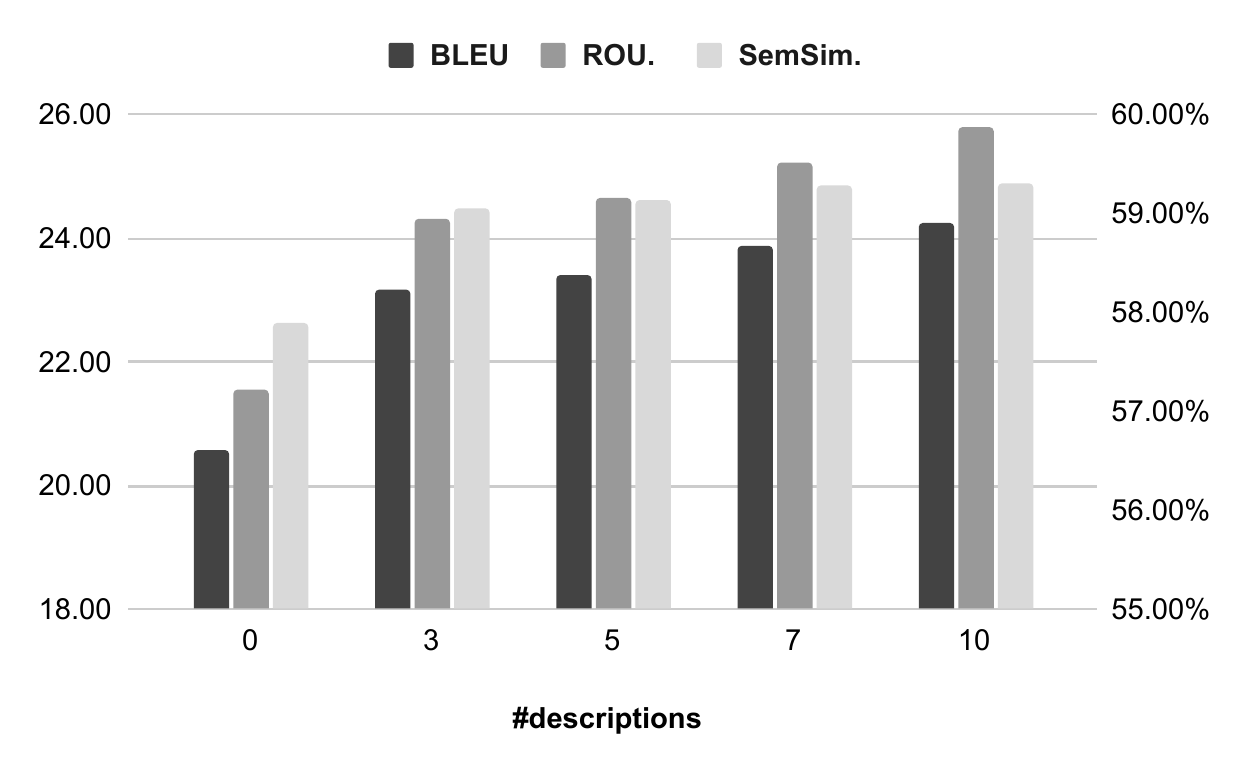}
    \caption{Impact of historical descriptions (left axis: \textit{BLEU} \& \textit{ROU}.; right axis: \textit{SemSim})}
    \label{fig:historical_descriptions}
\end{figure}

To investigate the impact of historical descriptions on \tool performance, we gradually increase the number of historical descriptions from 0 to 10.  
In this experiment,  we use CodeT5 as a translation model, using the $K$-means algorithm with \textit{Tf–idf} embedding to cluster description and apply it to the \dev setting. 
As shown in Figure~\ref{fig:historical_descriptions}, \tool can produce much better patch descriptions, up to about 18\%, when considering historical descriptions. This is because historical description helps the model capture a specific style or personalize convention of the author. Thus, the descriptions generated between \tool and the developer may be in the same style and expression. 
In addition, when increasing the number of historical descriptions from 1--10, the performance of the \tool is slightly improved from 23.18--24.23 in BLEU. These results indicate that although historical descriptions could be valuable for generating patch descriptions in \tool, considering more historical descriptions could not provide more new valuable information for generating patch descriptions. Thus, the list of historical descriptions should be shortened to maintain both the effectiveness and efficiency of \tool.

\subsection{Answering RQ4: Sensitivity Analysis}

\subsubsection{Impact of Training Dataset}

\begin{figure}
    \centering
    \includegraphics[width=\columnwidth]
    {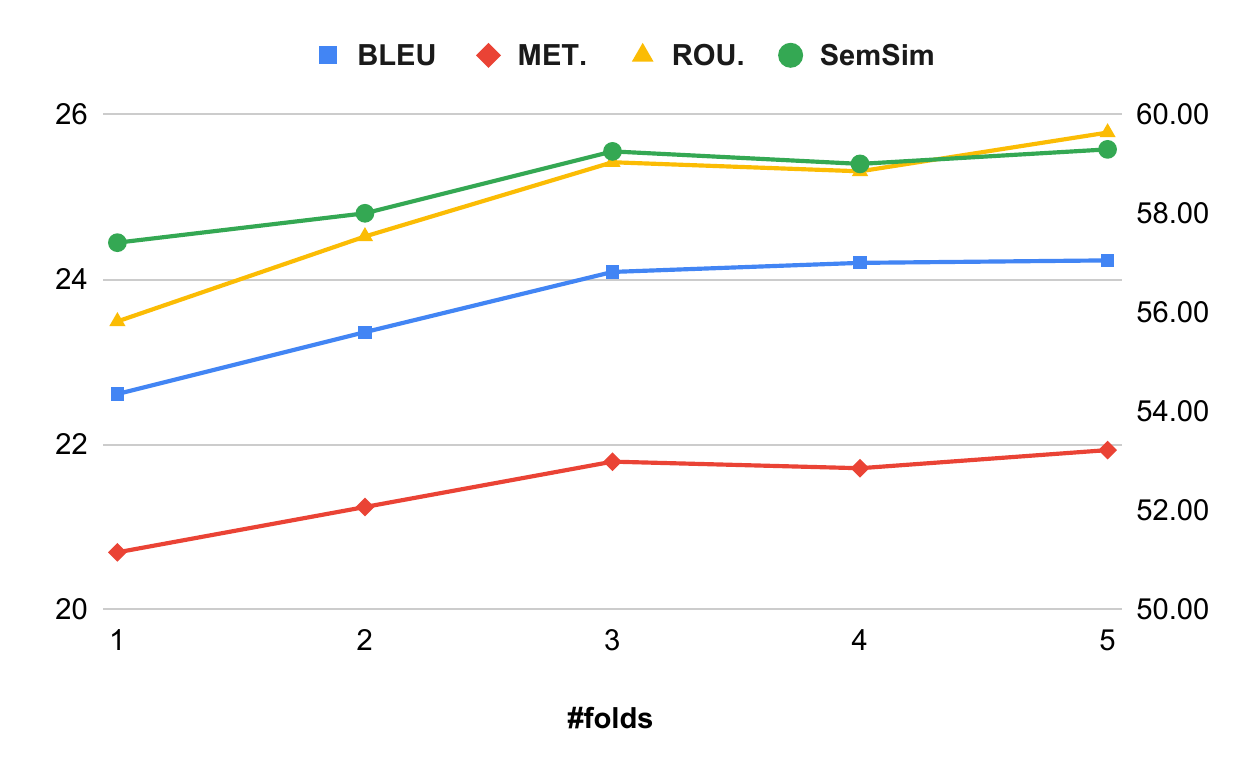}
    \caption{Impact of training data size (left axis: \textit{BLEU} \& \textit{MET}., and \textit{ROU}.; right axis: \textit{SemSim})}
    \label{tab:training_data}
\end{figure}

To measure the impact of training data size on \tool performance, we gradually increased the sizes of the training dataset by expanding the period for training data.
%
%
%
%
As shown in Figure~\ref{tab:training_data}, \tool demonstrates impressive efficiency in learning from training data. Although with more data, \tool can learn to understand patches and translate descriptions better, \tool attains 93\% of its full performance with only one-fifth of the training data (4.3K out of 21.8K instances). Moreover, \tool achieves 99\% of its performance with 60\% of the training data, significantly reducing the required training time. These results show \tool's capability to efficiently leverage a relatively smaller portion of the training data while still achieving high-performance levels. Such efficiency is crucial when computational resources or time constraints necessitate optimizing training efficiency. 


\subsubsection{Impact of Change Complexity}

\begin{figure}
    \centering
    \includegraphics[width=\columnwidth]
    {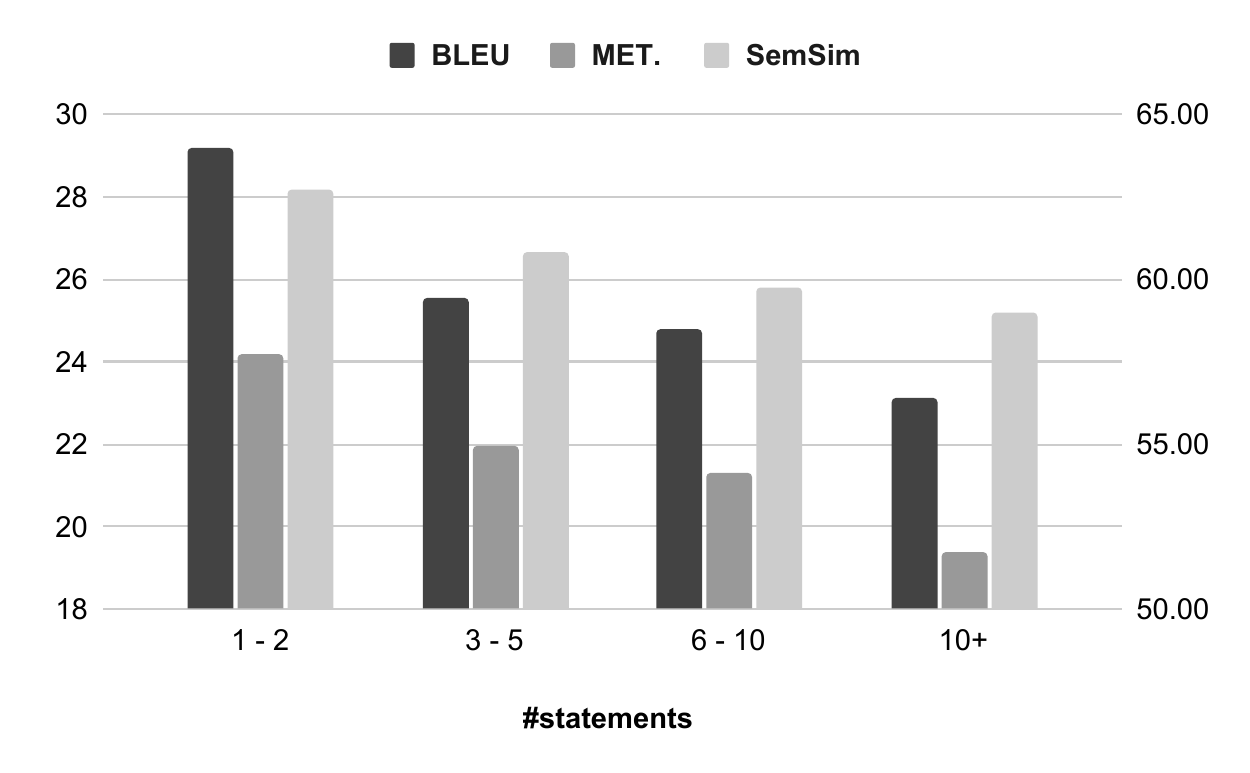}
    \caption{Performance of \tool in different change complexity levels by \#changed statements (left axis: \textit{BLEU} \& \textit{METEOR}; right axis: \textit{SemSim.})}
    \label{fig:change_complexity}
\end{figure}

To investigate the impact of change complexity on \tool's performance, we conducted evaluations using varying numbers of changed statements in the \dev setting. As shown in Figure~\ref{fig:change_complexity}, the performance of \tool slightly decreases by 6\% in \textit{Semantic Similarity} when increasing the number of changed statements.
This decline is anticipated because it is more challenging for \tool to capture the meaning of the patch having many changed statements.
However, the slight decrease implies that \tool remains a reliable choice for generating patch descriptions even in the patches involving more changes.

\subsection{Answering RQ5: Time Complexity}
In this work, all experiments were conducted on a GeForce RTX 4090 GPU with 24GB RAM. \tool required about 1.0 seconds to extract the contextual information for each case. Additionally, the patch clustering step in \tool costed 45 seconds to group more than 25K patches in the training dataset based on their description similarity. After that, \tool took approximately 10 hours for fine-tuning Code Llama 7B~\cite{codellama} as its translation model. On average, \tool required totally 0.59 seconds to generate a patch description. 


\subsection{Threats to Validility}

The main threats to the validity of our work consist of internal, construct, and external threats.

\textbf{Threats to internal validity} include the hyperparameter settings of all of the baseline models as well as the ones chosen for \tool. To reduce this threat, we systematically varied the setting of \tool to study its performance (Section~\ref{sec:results}) and reused the implementation with the same setting in the papers of the baseline approaches~\cite{nngen,coregen,come,race}.
A threat may come from the method used to extract patch-related code. To reduce this threat, we use Joern~\cite{joern} code analyzer, which is widely used in existing studies~\cite{velvet,vuldeeppeaker}. Another threat mainly lies in the correctness of the implementation of our approach. To reduce such a threat, we carefully reviewed our code and made it public~\cite{website} so that other researchers could double-check and reproduce our experiments.

\textbf{Threats to construct validity} relate to the suitability of our evaluation procedure. We used \textit{BLEU}, \textit{METEOR}, \textit{ROUGE}, and \textit{Semantic Similarity}. They are the widely-used evaluation measures for in the studies in code/change-text translation~\cite{codesum_paper,icse20,deep-comment-gen, Logentext,nngen,coregen,race,come,bug-explainer}. 
\hl{Another potential threat is the lack of human evaluation, which could provide a more comprehensive assessment of our approach. To address this, we plan to incorporate a human evaluation method in future work, where experienced software developers will review and rate the generated descriptions based on criteria such as accuracy, completeness, and usefulness.}
Additionally, evaluation in controlled environments may pose a threat. To mitigate this, we evaluated the approaches in two settings (\textit{dev. process} and \textit{cross-project}) and investigated the performance in various scenarios and settings.

\textbf{Threats to external validity} mainly lie in the selection of translation models used in our experiments. 
To mitigate this threat, we select the representative models that are well-known for NLP and SE tasks~\cite{codet5,codellama,vaswani2017attention}.
Another threat could be the potential limitations of using a set of patches and a limited set of keywords in investigating the ``what'', ``how'' and ``why'' aspects in Section~\ref{sec:example}. To mitigate this threat, we select a large set of 24.2K patches and perform a manual analysis of randomly sampled descriptions with multiple reviewers.
We will expand the dataset and alternative set of keywords to further mitigate this threat.
Moreover, our experiments are conducted on only the code changes of C/C++ projects. Thus, the results could not be generalized for other programming languages. In our future work, we plan to conduct more experiments to validate our results in other languages. 

\section{\hl{Related Work}}

\tool closely relates to the work on \textbf{automated commit message generation}~\cite{nngen,coregen,codedisum,race,come,history_paper,fira,vu2023context,kadel_2024,chatgpt_2024,critical_view_2024}.
%
NNGen~\cite{nngen}, an IR-based approach, leverages cosine similarity to identify the $k$ most similar commit diffs (represented using CC2Vec~\cite{cc2vec}) from a pre-existing dataset, eliminating the need for a separate training phase.
%
Coregen~\cite{coregen} adopts a purely Transformer-based architecture for learning representations specifically tailored to commit message generation.
%
RACE~\cite{race} integrates context-awareness into a retrieval-based deep learning approach for generating commit messages, effectively combining the strengths of language modeling and information retrieval techniques.
%
COME~\cite{come} employs a decision algorithm to combine retrieval strategies with translation-based methods, enabling the model to learn more contextually-aware representations of code changes.
%
%
While these approaches can generate commit messages, they often focus on explaining general code changes rather than the specific bug fixes addressed by the patch. 
These approaches implicitly learn semantic and conventional information during description generation. 
In contrast, our method is specialized for patch descriptions by explicitly providing semantic and conventional information, achieving demonstrably superior performance in Section~\ref{sec:results}.

Our work aligns with the broader area of \textbf{code-to-text translation}, where existing approaches translate code into natural language for various purposes like generating code comments, bug explanations, and source code summaries. 
Recent advancements heavily rely on learning-based methods~\cite{Logentext,deep-comment-gen,summarization_attention,code_summarization2,comment_gen,ir-guided-comment-gen, mastropaolo2021studying, gao2023code,icse20,comment_gen_2024,generate_log_2024}. 
Notably, Wei \etal~\cite{ir-guided-comment-gen} combine both
IR and NMT in comment generation. Mastropaolo \etal~\cite{mastropaolo2021studying} leverage the powerful Text-To-Text Transfer Transformer (T5) for various tasks including code comments.
Geng~\etal~\cite{comment_gen_2024} leverage large language models (LLMs) for code comment generation, addressing the limitation of traditional methods that generate only one comment per code snippet, which often fails to meet developers' diverse needs.
Gao \etal~\cite{gao2023code} introduce a structure-guided transformer for code summarization, integrating code structural features into the Transformer architecture. This involves embedding local symbolic details like code tokens and statements, along with global syntactic structures such as data flow graphs, into the self-attention module of the Transformer.
Xu~\etal~\cite{generate_log_2024} propose UniLog which is an automatic logging framework leveraging the in-context learning paradigm of large language models (LLMs).
Parvez~\etal~\cite{bug-explainer} propose Bugsplainer, a transformer-based generative model that generates natural language explanations for software bugs. Bugsplainer leverages structural information and buggy patterns from the source code to generate an explanation for a bug.
Importantly, these approaches describe the source code itself, while \tool focuses on the specific changes introduced in the code, particularly the bug fixes.
%
This complementary nature suggests that combining these approaches (including \tool) could provide developers with even more comprehensive explanations in natural language for various steps during the software development process.

Several \textbf{learning-based approaches} have been proposed for specific SE tasks such as code suggestion~\cite{icse20, naturalness,arist}, code generation~\cite{dong2023codep, zheng2023codegeex}, code translation~\cite{pan2024lost,zhang2023multilingual,structcoder}, test generation~\cite{liu2023fill, lan2024deeply}, bug detection~\cite{oppsla19, CodeJIT, qiu2024vulnerability,nguyen2024context}, bug fix identification~\cite{kse}, and program repair~\cite{xia2023automated,ruan2024timing}.
\section{Conclusion}

In conclusion, our proposed approach, \tool, addresses the challenges associated with manual patch description generation by framing it as a machine translation task. Leveraging explicit representations, historical context, and syntactic conventions, \tool demonstrates consistent superiority over existing methods in experimental evaluations. With improvements of up to 189\% in BLEU, 5.7X in Exact Match rate, and 154\% in Semantic Similarity, \tool showcases its efficacy in automating the generation of software patch descriptions. The dual objectives of maximizing similarity and accurately predicting affiliating groups contribute to the robust performance of \tool, establishing it as a valuable tool in the software development cycle.

\printcredits

\bibliographystyle{elsarticle-num}

\bibliography{ref}

\end{document}